\documentclass[aip,reprint,jcp,showpacs,preprintnumbers,amsmath,amssymb,superscriptaddress,a4paper]{revtex4-1}

\usepackage{graphicx}
\usepackage{bm}
\usepackage{units}

\begin{document}


\title{Direct Observation of High-Spin States in Manganese Dimer and Trimer Cations by X-ray Magnetic Circular Dichroism Spectroscopy in an Ion Trap}

\author{V.~Zamudio-Bayer}
	\affiliation{Physikalisches Institut, Universit\"{a}t Freiburg, Stefan-Meier-Stra{\ss}e 21, 79104 Freiburg, Germany}
	\affiliation{Institut f\"ur Methoden und Instrumentierung der Forschung mit Synchrotronstrahlung, Helmholtz-Zentrum Berlin f\"ur Materialien und Energie GmbH, Albert-Einstein-Stra{\ss}e 15, 12489 Berlin, Germany}

\author{K.~Hirsch}
	\altaffiliation[Present address: ]{Stanford Institute for Materials and Energy Sciences, SLAC National Accelerator Laboratory, 2575 Sand Hill Road, Menlo Park, California 94025, USA }
\author{A.~Langenberg}
	\altaffiliation[Present address: ]{Max Planck Institut f\"ur Plasmaphysik, Wendelsteinstra{\ss}e 1, 17491 Greifswald, Germany}
\author{M.~Kossick}
	\affiliation{Institut f\"ur Methoden und Instrumentierung der Forschung mit Synchrotronstrahlung, Helmholtz-Zentrum Berlin f\"ur Materialien und Energie GmbH, Albert-Einstein-Stra{\ss}e 15, 12489 Berlin, Germany}
	\affiliation{Institut f\"{u}r Optik und Atomare Physik, Technische Universit\"{a}t Berlin, Hardenbergstra{\ss}e 36, 10623 Berlin, Germany}

\author{A.~\L awicki}
	\affiliation{Institut f\"ur Methoden und Instrumentierung der Forschung mit Synchrotronstrahlung, Helmholtz-Zentrum Berlin f\"ur Materialien und Energie GmbH, Albert-Einstein-Stra{\ss}e 15, 12489 Berlin, Germany}

\author{A.~Terasaki}
	\affiliation{Cluster Research Laboratory, Toyota Technological Institute, 717-86 Futamata, Ichikawa, Chiba 272-0001, Japan}
	\affiliation{Department of Chemistry, Kyushu University, 6-10-1 Hakozaki, Higashi-ku, Fukuoka 812-8581, Japan}

\author{B.~v.~Issendorff}
	\affiliation{Physikalisches Institut, Universit\"{a}t Freiburg, Stefan-Meier-Stra{\ss}e 21, 79104 Freiburg, Germany}

\author{J.~T.~Lau}
	\email{tobias.lau@helmholtz-berlin.de}
	\affiliation{Institut f\"ur Methoden und Instrumentierung der Forschung mit Synchrotronstrahlung, Helmholtz-Zentrum Berlin f\"ur Materialien und Energie GmbH, Albert-Einstein-Stra{\ss}e 15, 12489 Berlin, Germany}

\date{\today}

\begin{abstract}
The electronic structure and magnetic moments of free Mn$_2^+$ and Mn$_3^+$ are characterized by $2p$ x-ray absorption and x-ray magnetic circular dichroism spectroscopy in a cryogenic ion trap that is coupled to a synchrotron radiation beamline. Our results show directly that localized magnetic moments of 5 $\mu_B$ are created by $3d^5 (^6\mathrm{S})$ states at each ionic core, which are coupled in parallel to form molecular high-spin states via indirect exchange that is mediated in both cases by a delocalized valence electron in a singly-occupied $4s$ derived orbital with an unpaired spin. This leads to total magnetic moments of 11 $\mu_B$ for Mn$_2^+$ and 16 $\mu_B$ for Mn$_3^+$, with no contribution of orbital angular momentum. 
\end{abstract}


\maketitle

\section*{Introduction}
Manganese is an element with peculiar electronic and magnetic properties. Of all $3d$ transition elements, the manganese atom carries the second largest magnetic moment of 5$\mu_B$ because of the high-spin $3d^5\ (^6\mathrm{S})$ subshell configuration, while bulk manganese has an unusual 58-atom unit cell with noncollinear antiferromagnetic order \cite{Hobbs03} below the N\'eel temperature. A similar complex behavior can also be found in manganese molecules and clusters \cite{Morse86,Alonso00}. 
Stern-Gerlach deflection studies of Mn$_n\ (n = 5 - 99)$ cluster beams \cite{Knickelbein01,Knickelbein04a} showed superparamagnetism or ferrimagnetism with average magnetic moments that oscillate between $0.4 - 1.7\ \mu_B$ per atom. Heisenberg behavior \cite{Negodaev08} was postulated for Mn$_2$ and a transition from ferro- to antiferromagnetism \cite{BobadovaParvanova05,Kabir06} with increasing cluster size, as well as noncollinear spin structure \cite{Pederson98,Longo08,Zeleny09}, was predicted. Combined photoelectron spectroscopy and density functional theory studies \cite{Gutsev08} found indications \cite{Jellinek06} for half-metallic Mn$_n$ clusters.
\newline
Turning to the smallest clusters and molecules, the electronic ground states of molecular manganese cations have been studied by matrix-isolation electron spin resonance spectroscopy \cite{VanZee81,Baumann83,VanZee88,Cheeseman90} and photodissociation spectroscopy \cite{Terasaki01,Terasaki03}. 
In this size range, Mn$_n^+$ cations are characterized by low dissociation energies \cite{Ervin83,Jarrold85,Terasaki03,Tono05} of 1.39 eV for Mn$_2^+$ and $0.83 \pm 0.05$ eV for Mn$_3^+$ that increase only slightly to $1.06 \pm 0.03$ eV for Mn$_7^+$.  
For Mn$_2^+$, the experimental studies agree on a $^{12}\Sigma_g^+$ ground state as do most theoretical studies  \cite{Bauschlicher89,Nayak98,Desmarais00,Wang05} even though a $^{10}\Pi_u$ ground state \cite{Gutsev03a} is also considered. 
For Mn$_3^+$ a $^5B_2$ ground state \cite{Gutsev06} is predicted by theory but photodissociation spectroscopy \cite{Terasaki03} favors a $^{17}B_2$ high-spin ground state. Local high-spin states \cite{Lau09b,Hirsch12b} of Mn$_2^+$ and Mn$_3^+$ are also inferred from x-ray absorption spectroscopy. 
This illustrates the need for a direct experimental probe for the spin states of size-selected free molecular ions \cite{ZamudioBayer15}.
\newline
One interesting aspect of the interplay of magnetism and chemical bonding in manganese is that large magnetic moments could in principle be obtained if the atomic Mn $3d^5\ (^6\mathrm{S})$ high-spin state could be preserved in larger manganese entities and if parallel spin alignment could be achieved by long range ferromagnetic interaction.
Here we show experimentally that Mn$_2^+$ and Mn$_3^+$ are characterized by fully occupied majority spin states and local $3d^5\ (^6\mathrm{S})$ high spin terms that couple to 11 $\mu_B$ and 16  $\mu_B$, respectively. Non-collinear spin arrangements can be ruled out for the smallest molecular cations. 

\section*{Experimental and Computational Details}

\subsection*{Sample Preparation}
Mn$_2^+$ and Mn$_3^+$ were prepared in situ in a cluster ion beam apparatus \cite{Lau08,Hirsch09,Niemeyer12} by direct-current magnetron sputtering of a high-purity manganese target (99.95 \%, Lesker) in a mixed (approx.\ 5:1 volume flow ratio) helium-argon (99.9999 \%) atmosphere of 0.1 - 1 mbar at liquid nitrogen temperature. The magnetron discharge creates neutral and ionic species that grow by gas aggregation. A distribution of Mn$_n^+$ ions was extracted from the ion source and guided through differential pumping stages into a radio-frequency quadrupole mass filter (Extrel) to select either Mn$_2^+$ or Mn$_3^+$ parent ions, which were then stored in a liquid-helium-cooled linear quadrupole ion trap \cite{Niemeyer12,ZamudioBayer13,Langenberg14,ZamudioBayer15,Hirsch15a} filled with $10^{-4} - 10^{-3}$ mbar high purity ($> 99.9999$ \%) helium buffer gas. The number density of helium atoms in the ion trap is $\approx 7 \cdot 10^{13} - 9 \cdot 10^{14}$ cm$^{-3}$ at our experimental parameters. 
Under these conditions, vibrations and rotations are thermalized to equilibrium on a time scale of micro to milliseconds \cite{Gerlich95,Gerlich09,Otto13,Hansen14,Boyarkin14}. The ion trap was continuously filled with parent ions to the space charge limit. Typical storage times of the parent ions in the ion trap were $1 - 10$ s. This excludes the possibility of trapping metastable configurations \cite{Hirsch12a}. The purity of the parent ions in the ion trap was verified by reflectron time-of-flight mass spectrometry. 
The homogeneous static magnetic field of a superconducting solenoid \cite{Terasaki07} (JASTEC) that surrounds the ion trap vacuum chamber was used to magnetize the Mn$_2^+$ and Mn$_3^+$ samples. The inhomogeneity of the applied magnetic field is $\le 1$ \% over the entire ion trap volume. 

\subsection*{Spectroscopic Technique}
X-ray absorption and x-ray magnetic circular dichroism (XMCD) spectroscopy at the manganese $L_{2,3}$ edges of the Mn$_2^+$ and Mn$_3^+$ parent ions was performed inside the ion trap in ion yield mode by monitoring the intensity of Mn$^{2+}$ product ions in both cases. These product ions are generated by dissociation of highly excited intermediates that result from x-ray absorption that is followed by Auger decay of the $2p$ core-excited state of the parent ion. Parent and product ion bunches (a small fraction of the trap filling) were extracted at a rate of $\approx 0.3$ kHz from the ion trap and guided into the acceleration region of the reflectron time-of-flight mass spectrometer for detection. The incident photon energy was scanned across the manganese $L_{2,3}$ absorption edges from $610 - 690$ eV in $250 - 500$ meV steps with 625 meV photon energy resolution. At every photon energy step, the sample was irradiated with monochromatized x-rays  for 8 s and the product ion intensity was recorded in a photoionization mass spectrum. For XMCD spectroscopy, a static magnetic field with $\mu_0$H = 5 T was applied along the ion trap axis, with parallel or antiparallel orientation to the photon helicity of the incoming elliptically-polarized soft-x-ray beam. 
The difference of the spectra that are recorded for negative and positive helicity of the x-ray photons gives the XMCD spectrum, and the average is the isotropic x-ray absorption spectrum.
All spectra were normalized to incoming photon flux, detected by a GaAsP photo diode, and were corrected for the 90 \% polarization degree of the elliptically polarized soft x-ray photons. The experiments were carried out at beamlines UE52-SGM and UE52-PGM of the BESSY II synchrotron radiation facility at Helmholtz-Zentrum Berlin. 

\subsection*{Atomic Hartree-Fock Calculations}
To analyze the experimental data, we have calculated the x-ray absorption and XMCD spectra of Mn$^+$ for the [Ar]\,$3d^5\,4s^1$ configuration in the $^7$S ground state and $^5$S excited state term. Dipole accessible $2p^5\,3d^6\,4s^1$ and $2p^5\,3d^5\,4s^2$ final state configurations were taken into account. 
These Hartree-Fock calculations were performed with the {\sc{cowan code}} \cite{Cowan68} as implemented in {\sc{missing}} \cite{MISSING}. In our calculations the usual scaling-down of the Coulomb and exchange interaction parameters to 85 \% of the \emph{ab initio} values was applied in order to account for intra-atomic relaxation effects \cite{Cowan68}. The calculated spectra were convoluted with a lifetime (Lorentz) broadening of 0.1 eV at the $L_3$ and 0.2 eV at the $L_2$ resonance, and an instrument (Gaussian) broadening with 0.25 eV full width at half maximum is applied to match the experimental photon energy resolution. The increased broadening at the $L_2$ resonance is due to the reduced lifetime of the $2p_{\nicefrac{1}{2}}$ core hole \cite{vanderLaan91}. 
The calculated spectra were redshifted by 2.57 eV in order to match the experimental excitation energy, and the \emph{ab initio} $2p$ spin-orbit splitting parameter $\zeta$ is 6.74 eV. Direct $2p$ photoionization was not included in the Hartree-Fock calculation. This causes the offset of the experimental x-ray absorption spectrum (cf.\ Fig.\ \ref{fig:XSpectra}) at higher excitation energies, but has no effect on the XMCD spectrum.

\section*{Results}

\subsection*{Atomic Localization of 3d Electrons in Mn$_2^+$ and Mn$_3^+$}
In Fig.\ \ref{fig:XSpectra} the experimental $L_{2,3}$ x-ray absorption and XMCD spectra of Mn$^+_2$ and Mn$_3^+$ are shown along with the corresponding theoretical spectra of Mn$^+$ in its [Ar]\,$3d^5\,4s^1$\ $^7$S ground state configuration. All spectra were normalized to the integrated signal of resonant $2p \rightarrow 3d$ transitions, i.e., to the number of unoccupied $3d$ states. This normalization and the good agreement of the spectral fingerprints allows us to directly compare theoretical and experimental XMCD signals in order to obtain information on the electronic ground states of Mn$_2^+$ and Mn$_3^+$.
\newline
For the Mn$_2^+$ and Mn$_3^+$ molecular ions, the respective XAS and XMCD spectra  in Fig.\ \ref{fig:XSpectra} are identical in shape to the calculated spectrum \cite{Lau09b,Hirsch12a,Hirsch12b} of atomic Mn$^+$ in its ground state configuration. This immediately shows that the experimental spectra originate from an unperturbed atomic $3d^5$ $(^6\mathrm{S})$ electronic configuration of the $3d$ subshell, i.e., the $3d$ electrons form local high-spin states but do not or only very weakly participate in bonding \cite{Lau09b,Hirsch12a,Hirsch12b,ZamudioBayer15} in Mn$_2^+$ and Mn$_3^+$. 
The fact that the $3d$ orbitals remain atomically localized \cite{Lau09b,Hirsch12a,Hirsch12b,ZamudioBayer15} has implications for the geometric structure. For unperturbed $3d$ orbitals, the overlap of $3d$ electrons at different nuclei must be very weak. 
A rough estimate of the interatomic distance \cite{ZamudioBayer15} in Mn$_2^+$ and Mn$_3^+$ can therefore be made with the atomic manganese $3d$ radial distribution function, from which it can be seen that the radial $3d$ electron density decreases to less than 1 \% of its maximum at $r \geq 1.3$ {\AA}, leading to a corresponding equilibrium distance for pure $4s\sigma$ bonding $r_e(\text{Mn}^+_2) \geq 2.6$ {\AA}. This estimate is in good agreement with calculated values \cite{Nesbet64,Bauschlicher89,Terasaki01,Wang05} of $r_e \approx 2.9 - 3.0$ {\AA} for Mn$_2^+$ and Mn$_3^+$ high-spin states. 
Thus the equilibrium distance in Mn$_2^+$ is significantly larger than the nearest neighbor distance in bulk manganese \cite{Hobbs03} of 2.24 {\AA} but also larger than the typical bond distances of diatomic transition metal cations, e.g.,\ the experimental values \cite{Asher94,Yang00} of $r_e(\text{V}^+_2) = 1.73$ {\AA} and $r_e(\text{Ni}^+_2) = 2.22$ {\AA}, where $3d$ orbitals participate in molecular bonding.

\subsection*{High-Spin Ground State of Mn$_2^+$}
For the lowest energy dissociation limit of Mn$_2^+$ into Mn $3d^5\,4s^2\ ^6$S and Mn$^+$ $3d^5\,4s^1\ ^7$S, the localized $3d^5\ (^6\mathrm{S})$ high-spin states at the manganese cores and the delocalized single $4s$ derived spin could couple to states with a total spin \cite{Wigner28} $S$ of $11 \ge 2S \ge 1$ in Mn$_2^+$. In the following treatment we will assume for simplicity that the localized $3d^5\ (^6\mathrm{S})$ states first couple to a total $3d$ spin $S_{3d}$ and then with the single unpaired $4s$ derived spin to give a total spin $S$. This should be a good approximation of the real angular momentum coupling. 
\begin{figure}
	\includegraphics{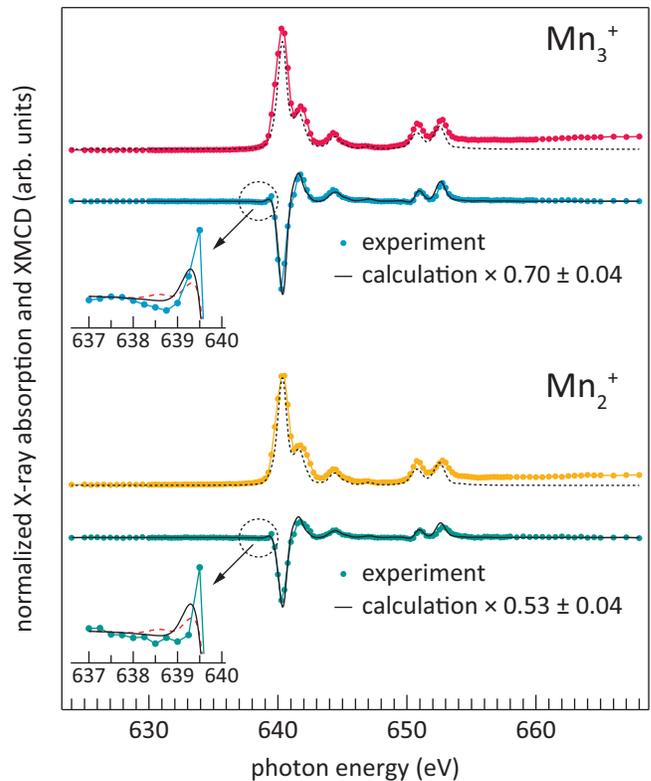}
	\caption{\label{fig:XSpectra} Experimental $2p$ x-ray absorption and XMCD spectra (bullets) of Mn$_2^+$ and Mn$_3^+$ along with x-ray absorption (broken line) and XMCD (solid line) from an atomic Mn$^+$ $3d^5\,4s^1\ ^7$S Hartree-Fock calculation \cite{Cowan68,MISSING}. Calculated XMCD spectra are scaled to match the experimental amplitude to give the experimental magnetization. Inset: detailed pre-edge region of experimental and theoretical XMCD spectra. The strong overshoot at 639.5 eV that is observed in the experimental spectra agrees better with the overshoot that is obtained in the calculated spectrum for the initial [Ar]\,$3d^5\, 4s^1$ configuration in the $^7$S term (solid line) than in the $^5$S term (dashed line) of Mn$^+$. This indicates parallel $3d - 4s$ coupling.}
\end{figure}
\newline
Applied to the experimental spectrum, the orbital angular momentum sum rule of XMCD \cite{Thole92,Carra93} yields a molecular orbital magnetic moment of $\mu_L = (0.1 \pm 0.4)$ $\mu_B$ for Mn$_2^+$ and clearly indicates a $\Sigma$ ground state. The total angular momentum $J$ of Mn$_2^+$ thus is exclusively spin. 
\newline
Because of the identical atomic $3d^5$ signature in the x-ray absorption and XMCD spectra, the experimental $3d$ spin magnetization can be obtained by fitting the calculated Mn$^+$ XMCD signal to the experimental Mn$_2^+$ spectrum \cite{ZamudioBayer15,Hirsch15a} after normalization. 
This circumvents the empiric correction \cite{OBrien94b,Piamonteze09} that would be required to the 
XMCD spin sum rule \cite{Carra93} for manganese. 
\newline
From the fit of the calculated XMCD of Mn$^+$ to the experimental spectrum of Mn$_2^+$ in Fig.\ \ref{fig:XSpectra}, a total $3d$ spin magnetization of $(0.53 \pm 0.04) \cdot 2\cdot 5\ \mu_B\ = (5.3 \pm 0.4)\ \mu_B$ is determined. 
This rules out all states with $2S_{3d} \le 4$ and leaves only six possible states with $2S_{3d} = 6\ (2S = 5, 7)$, $2S_{3d} = 8\ (2S = 7, 9)$, and $2S_{3d} = 10\ (2S = 9, 11)$. 
The temperature at which $S_{3d}$ of these states would reach a magnetization of $(5.3 \pm 0.4)\ \mu_B$ at $\mu_0H = 5$ T is given by the Brillouin function for the total spin $S$. 
An ion temperature of $\le 7$ K that would be necessary for the $2S_{3d} = 6$ states can be ruled out because the experiment was performed at a temperature of the ion trap of $8 \pm 1$ K and radio-frequency heating of the ions is inevitable at our conditions \cite{Niemeyer12,Langenberg14,ZamudioBayer15}. 
For a $3d$ magnetization of $(5.3 \pm 0.4)\ \mu_B$ in the two $2S_{3d} = 8\ (2S = 7, 9)$ states, an ion temperature of $11 \pm 2$ K and $14 \pm 2$ K would be required, which would correspond to a radio-frequency heating of $3 \pm 2$ K and $6 \pm 2$ K, respectively. As will be shown below, radio-frequency heating at the conditions of our experiment is $\ge 10 \pm 2$ K for Mn$_2^+$, which rules out both states.
\newline
The remaining $2S_{3d} = 10\ (2S = 9, 11)$ states with fully parallel alignment of the $3d$ states but antiparallel $(2S = 9)$ or parallel $(2S = 11)$ alignment of the $4s$ spin would correspond to ion temperatures of $20 \pm 2$ K and $23 \pm 2$ K, respectively, and cannot be distinguished by consideration of the very similar radio frequency heating of $12 \pm 2$ K and $15 \pm 2$ K. 
However, the $2S_{3d} = 10\ (2S = 9)$ state with antiparallel coupling of the $4s$ derived electron spin should be about 1 eV higher in energy than the $2S_{3d} = 10\ (2S = 11)$ state because of the strong intra-atomic $3d - 4s$ exchange coupling in manganese that leads to parallel spin coupling and favors  the $3d^5\,4s^1\ ^7$S term over the $3d^5\,(^6\mathrm{S})\,4s^1\ ^5$S term in the ground state of free Mn$^+$ ions \cite{Sugar85} by 1.17 eV. This argument is very similar to the case of the maximum spin ground state \cite{ZamudioBayer15} of Cr$_2^+$. 
\newline
Parallel $3d - 4s \sigma$ spin alignment is also indicated at the onset of the $L_3$ line of the experimental XMCD spectrum as shown in detail as an inset to Fig.\ \ref{fig:XSpectra}. Here the experimental spectra of Mn$_2^+$ and Mn$_3^+$ are compared to calculated spectra for $^7$S and $^5$S terms, i.e., for parallel and antiparallel alignment of the $4s$ derived spin. As can be seen, the calculated XMCD spectra differ significantly in the intensity of the overshoot at 639.25 eV, and by the sign of the signal at 638.5 eV. The strong overshoot at 639.5 eV and the dip at 638.75 eV in the experiment agree better with the calculated spectrum for parallel than for antiparallel alignment. 
In addition to the energy consideration above, this is an experimental indication of parallel alignment of the $4s$ derived spin and leaves only $2S_{3d} = 10\ (2S = 11)$ for the spin in the ground state of Mn$_2^+$. 
A similar signature of parallel $3d - 4s \sigma$ spin coupling has also been observed in the case of Cr$_2^+$ where it appears not only in XMCD but also in the x-ray absorption spectrum as a well separated line at photon energies below the $L_3$ main line \cite{Lau09b,ZamudioBayer15}. This prepeak at the onset of $2p$ excitation of the free molecular ion is a sensitive measure of parallel $3d - 4s$ spin alignment that vanishes upon cluster deposition \cite{Lau09b,ZamudioBayer15,Lau00b,Lau05,Reif05}. 
\newline
In conclusion, our results from XMCD spectroscopy provide the first direct experimental evidence for the $^{12}\Sigma^+_g$ ground state of free $\mathrm{Mn}_2^+$ ions, in agreement with electron spin resonance spectroscopy of $\mathrm{Mn}_2^+$ isolated in inert gas matrices~\cite{VanZee88,Cheeseman90} and in agreement with indirect evidence from photodissociation spectroscopy of free $\mathrm{Mn}_2^+$ ions~\cite{Terasaki01}. 

\subsection*{High-Spin Ground State of Mn$_3^+$}
The analysis of the spin and orbital angular momentum in the ground state of Mn$_3^+$ follows the same reasoning as for Mn$_2^+$ above. 
In Mn$_3^+$, the localized $3d^5\ (^6\mathrm{S})$ configurations of the $3d$ subshells can couple with the $4s$ derived spin to a total spin $S$ of $16 \ge 2S \ge 0$ as given by the 2 Mn ($3d^5\,4s^2\ ^6$S) + Mn$^+$ ($3d^5\,4s^1\ ^7$S) lowest energy dissociation limit. 
We will again assume for simplicity that the localized $3d^5\ (^6\mathrm{S})$ states in Mn$_3^+$ first couple to a total $3d$ spin $S_{3d}$ and then with the single unpaired $4s$ derived spin to give a total spin $S$.
\newline
Again, the orbital angular momentum sum rule of XMCD \cite{Thole92,Carra93} returns a molecular orbital magnetic moment of $\mu_L = (0.06 \pm 0.48)$ $\mu_B$ for Mn$_3^+$, i.e., no orbit contribution to the total angular momentum $J$ in agreement with half-filled $3d^5$ subshells localized at each of the three manganese cores. 
\newline
As for Mn$_2^+$, the total magnetization of the $3d$ spins in Mn$_3^+$ is determined from a fit of the calculated Mn$^+$ XMCD signal to the experimental XMCD spectrum as shown in Fig.\ \ref{fig:XSpectra}. This procedure gives $\left(0.70 \pm 0.04 \right) \cdot 3 \cdot 5\ \mu_B = (10.5 \pm 0.6) \ \mu_B$ as the total $3d$ magnetization and excludes all states with $2S_{3d} \le 9$. 
Out of the remaining states, those with $2S_{3d} = 11\ (2S = 10, 12)$ and $2S_{3d} = 13\ (2S = 12, 14)$ can be ruled out because these would reach a $3d$ magnetization of $(10.5 \pm 0.6) \ \mu_B$ at $\mu_0H = 5$ T only at ion temperatures of $\le 7$ K and $12 \pm 2$ K, respectively, according to the Brillouin function. These ion temperatures are incompatible with an ion trap temperature of $10 \pm 1$ K and radio-frequency heating that is inevitably present in the experiment. 
This result indicates fully parallel spin coupling of the localized $3d^5$ subshells to the maximum possible value of $2S_{3d} = 15$ in Mn$_3^+$. In principle, two states with a total spin of $2S = 14$ and $2S = 16$ could be formed by antiparallel or parallel coupling of the $4s$ derived spin with the $3d$ spin $S_{3d}$.
Analogous to the case of Mn$_2^+$, antiparallel coupling can be ruled out from energy considerations and from the experimental XMCD spectrum, shown in detail as an inset to Fig.\ \ref{fig:XSpectra}. Again, the state with antiparallel coupling of the $4s$ derived spin to $S_{3d}$ should be $\approx 1.17$ eV higher in energy \cite{Sugar85} because of the strong intra-atomic $3d - 4s$ exchange interaction that the $4s$ derived electron experiences with the localized $3d$ electrons. 
Furthermore, the calculated spectrum for the $^7$S ground state of Mn$^+$ with parallel coupling fits the details of the experimental spectrum at the onset of $2p$ excitation better than the $^5$S excited state with antiparallel $4s - 3d$ coupling. The case is is even clearer than for Mn$_2^+$ because of the better signal-to-noise ratio.
\newline
In summary, the total magnetic moment of Mn$_3^+$ is equal to $16\ \mu_B$ and is purely determined by the electron spin of the molecule. This is in agreement with results from photodissociation spectroscopy \cite{Terasaki03} of Mn$_3^+$. 
\newline
A total $3d$ magnetization of $(10.5 \pm 0.6) \ \mu_B$ at $\mu_0H = 5$ T for the $2S = 16\ (2S_{3d} = 15)$ state of Mn$_3^+$ corresponds to an ion temperature of $20 \pm 2$ K and thus to a radio-frequency heating of $10 \pm 2$ K at our experimental conditions. Since radio-frequency heating is more pronounced for Mn$_2^+$ than for Mn$_3^+$ because of the lighter mass at otherwise identical conditions, this result serves as an \emph{ex post} justification of the anticipated strong radio frequency heating of Mn$_2^+$ that was made above.

\section*{Discussion}
Our experimental results show that Mn$_2^+$ and Mn$_3^+$, similar to Cr$_2^+$ (Ref. \onlinecite{ZamudioBayer15}), possess electronic ground states with maximum $3d$ spin magnetic moments of 5 $\mu_B$ per atom and fully occupied $4s$ and $3d$  majority spin states. 
This confirms previous experimental results \cite{VanZee88,Cheeseman90,Terasaki01,Terasaki03,Lau09b,Hirsch12b} on free and matrix-isolated Mn$_2^+$ and Mn$_3^+$ but contradicts the theoretical predictions of a $^5B_2$ ground state \cite{Gutsev06} for Mn$_3^+$ and a $^{10}\Pi_u$ ground state \cite{Gutsev03a} for Mn$_2^+$.
\newline
The mechanism that mediates parallel spin alignment of the $3d^5\ (^6\mathrm{S})$ high spin configurations located at the individual nuclei in Mn$_2^+$ and Mn$_3^+$ is indirect (double) exchange coupling \cite{Zener51a,Anderson55} via the spin of a single unpaired $4s$ derived electron. This is identical in both molecular ions. 
The reason for this indirect exchange coupling is the strong intra-atomic $3d - 4s$ exchange interaction in manganese that leads to an energy difference of 1.17 eV between the $3d^5\,(^6\mathrm{S})\,4s^1\ ^7$S ground state of Mn$^+$ with parallel alignment of the unpaired $3d$ and $4s$ spins, and the $3d^5\,(^6\mathrm{S})\,4s^1\ ^5$S first excited state where the alignment is antiparallel \cite{Sugar85}. 
This intra-atomic exchange interaction favors a ferromagnetic coupling of the localized $3d$ shells in Mn$_2^+$ and Mn$_3^+$ because the unpaired $4s$ derived electron can only delocalize over the entire molecular ion with its spin aligned parallel to the $3d$ spins at the atomic sites. This is analogous to spin coupling \cite{ZamudioBayer15} in the ground state of Cr$_2^+$.
Indirect exchange coupling \cite{Zener51a} was first proposed for Mn$_2^+$ by \textcite{Bauschlicher89} who discussed how competing effects of intra-atomic $3d - 4s$ and interatomic $3d - 3d$ exchange coupling in Mn$^+_2$ would result in parallel coupling of the spins of the unpaired $4s \sigma$ electron and both localized $3d$ subshells in the ground state. A similar approach by \textcite{Wang05} regards Mn$_2^+$ as a mixed valence system and leads to identical results. 
\newline
One important requirement for indirect or double exchange \cite{Zener51a} is atomic-like spin correlation in the open $3d$ shell, which is not shown easily in general. Here, our experimental results demonstrate that this atomic-like spin correlation is clearly valid for Mn$_2^+$ and Mn$_3^+$ as for the case of Cr$_2^+$ from the atomic-like x-ray absorption and XMCD spectra \cite{Lau09b,Hirsch12a,Hirsch12b,ZamudioBayer15} that are very sensitive to spin correlations in the initial and final states. 
\newline
Finally, it is known that the XMCD spin sum rule \cite{Carra93} cannot be applied without empiric correction to manganese \cite{Duerr97,Piamonteze09} because $L_3$ and $L_2$ edges are not strictly separable as required. This leads to an underestimation of the spin magnetization. A comparison of the magnetization of Mn$_2^+$ and Mn$_3^+$ as obtained above to the result that would be given by application of the XMCD spin sum rule to the experimental spectra yields experimental correction factors to the spin sum rule of $1.7\pm0.3$ for Mn$_2^+$ and $1.5\pm0.1$ for Mn$_3^+$. These values agree well with results of 1.5 and 1.47 that are obtained by Hartree-Fock or charge-transfer multiplet calculations \cite{Duerr97,Piamonteze09}.

\section*{Conclusion}
In conclusion, x-ray absorption and x-ray magnetic circular dichroism spectroscopy of free Mn$_2^+$ and Mn$_3^+$ give direct experimental evidence of molecular high spin states. For both molecular ions we find a maximum $3d$ spin ground state with atomically localized $3d$ electrons and fully parallel alignment of all $3d$ electron spins, which leads to the largest possible magnetic spin moments of 11 $\mu_B$ and 16 $\mu_B$, respectively. We do not see any evidence for noncollinear or canted spin arrangements in Mn$_2^+$ and Mn$_3^+$.
\newline
We find that the same mechanism is responsible for the electronic structure and chemical bonding in Mn$_2^+$ and Mn$_3^+$. These molecular ions are characterized by localized $3d^5\ (^6\mathrm{S})$ high-spin configurations of half-filled $3d$ subshells, and a single delocalized $4s$ derived electron that mediates a strong and ferromagnetic indirect exchange coupling, analogous to the case of Cr$_2^+$.
From the observed localized atomic magnetic moment and $3d^5$ configuration we deduce an experimental lower limit on the interatomic distances of $r_e \geq 2.6$ {\AA} in agreement with theoretical predictions for the high spin states. There is no orbital contribution to the total angular momentum and thus no coupling of the spin to the molecular axis or frame to first order. 

\begin{acknowledgments}
Beam time for this project was granted at BESSY II beamlines UE52-SGM and UE52-PGM, operated by Helmholtz-Zentrum Berlin. Skillful technical assistance by Thomas Blume is gratefully acknowledged. This project was partially funded by the German Federal Ministry of Education and Research (BMBF) through grant BMBF-05K13Vf2. The superconducting solenoid was kindly provided by Toyota Technological Institute. AT acknowledges financial support by Genesis Research Institute, Inc. BvI acknowledges travel support by Helmholtz-Zentrum Berlin. 
\end{acknowledgments}


%

\end{document}